\documentclass{article}
\usepackage{amssymb}
\usepackage{amsfonts}
\usepackage{amsmath}

\begin{document}

\begin{center}
{\Large Codes over a subset of Octonion Integers }%
\begin{equation*}
\end{equation*}

Cristina Flaut

\begin{equation*}
\end{equation*}
\end{center}

\textbf{Abstract. }{\small In this paper we define codes over some Octonion
integers. We prove that in some conditions these codes can correct up to two
errors for a transmitted vector and the code rate of the codes is grater
than the code rate of the codes defined on some subset of Quaternion
integers.}%
\begin{equation*}
\end{equation*}


\begin{equation*}
\end{equation*}

\textbf{0. Preliminaries}%
\begin{equation*}
\end{equation*}

Coding Theory is a mathematical domain with many applications in Information
Theory. Various type of codes with their parameters were intensively
studied. As one of the important parameters of a code, the distance
associated (as Hamming, Lee, Mannheim, etc.) was also studied for many types
of codes, finding values for the minimum or maximum values for such a
distances ( see [Ne, In,Fa, El, Pa; 01], [Sa, Na, Re, Sa; 13] ). \newline
\qquad One of the type of codes with a great development in the last years
are Integer Codes.\ Integer codes are codes defined over finite rings of
integers modulo $m,m\in \mathbb{Z}$ and have some advantages over the
traditional block codes. One of them is that integer codes are capable to
correct limited number of error patterns which occur most frequently while
the conventional codes intend to correct all possible error patterns.
Integer codes have low encoding and decoding complexity and are suitable for
application in real communication systems ( see [Ko, Mo, Ii, Ha, Ma;10]).
There are some other codes similar to the integer codes, as for example
codes over Gaussian integers ([Hu; 94], [Gh, Fr; 10], [Ne, In,Fa, El, Pa;
01], [Ri; 95]), codes over Eisenstein--Jacobi integers ([Ne, In,Fa, El, Pa;
01]), a class of error correcting codes based on a generalized Lee distance
([Ni, Hi; 08]), codes over Hurwitz integers ([Gu; 13]), which are
intensively studied in the last time. \qquad

QAM \ is used in many digital data radio communications and data
communications applications. The most common errors which appear are those
which change a point with its nearest neighbor. The Hamming distance and the
Lee distance are not capable to correct these errors in a QAM signal. To
improve this situation, in [Hu; 94], Huber constructed codes over Gaussian
integers with a new distance, called Mannheim distance. He proved that these
codes are able to correct Mannheim error of weight $1$ and used this new
distance to find properties of these codes (see (see [Mo, Ha, Ko; 04]) for
other details). Nevertheless, in [Ni, Hi; 08] the authors introduced a new
distance which generalized the Lee distance and constructed codes capable to
correct errors of generalized Lee weight one or two.

In [Gu; 13], the author generalized some results from [Ne, In,Fa, El, Pa;
01] constructing codes over Hurwitz integers.

Due to the structure of the real Octonion algebra, a nonassociative and a
noncommutative algebra, in the presented paper, we generalize these results
to a special subset of Octonion integers, comparing with the some results
obtained until now. 
\begin{equation*}
\end{equation*}

\textbf{1. Introduction}%
\begin{equation*}
\end{equation*}

The octonion division algebra over $\mathbb{R},$ denoted by $\mathbb{O}%
\left( \mathbb{R}\right) ,$ is a nonassociative unital algebra. This algebra
is \textit{power-associative} (i.\thinspace e. the subalgebra $<x>$ of $%
\mathbb{O}\left( \mathbb{R}\right) $, generated by any element $x\in \mathbb{%
O}\left( \mathbb{R}\right) $, is associative), \textit{flexible}
(i.\thinspace e. $x(yx)=(xy)x=xyx$, for all $x,y\in \mathbb{O}\left( \mathbb{%
R}\right) )$ and has the following properties:

1) $\mathbb{O}\left( \mathbb{R}\right) $ is a free $\mathbb{R}-$module with
the basis $\{1,e_{2},e_{3},e_{4},e_{5},e_{6},e_{7},e_{8}\}.$

2) $1$ is the unity in $\mathbb{O}\left( \mathbb{R}\right) .$

3) $e_{2}^{2}=e_{3}^{2}=e_{4}^{2}=e_{5}^{2}=e_{6}^{2}=e_{7}^{2}=e_{8}^{2}=-1$
and $e_{i}e_{j}=-e_{j}e_{i}=e_{k},i\neq j,i,j\in \{2,...,8\},$ where $%
k=i\otimes j,$ where $\otimes $ is "x-or" for $i,j$ $\ $wrote in the decimal
basis (see [Ba; 09]).

If $%
x=x_{1}+x_{2}e_{2}+x_{3}e_{3}+x_{4}e_{4}+x_{5}e_{5}+x_{6}e_{6}+x_{7}e_{7}+x_{8}e_{8}\in 
\mathbb{O}\left( \mathbb{R}\right) ,$ then its \textit{conjugate} is the
octonion $\overline{x}%
=x_{1}-x_{2}e_{2}+x_{3}e_{3}+x_{4}e_{4}+x_{5}e_{5}+x_{6}e_{6}+x_{7}e_{7}+x_{8}e_{8} 
$ and the \textit{norm} of the octonion $x$ is 
\begin{equation*}
N\left( x\right) =x\overline{x}=\overline{x}%
x=x_{1}^{2}+x_{2}^{2}+x_{3}^{2}+x_{4}^{2}+x_{5}^{2}+x_{6}^{2}+x_{7}^{2}+x_{8}^{2}.
\end{equation*}%
\textbf{Remark 1.1.} The octonionic norm $N$ is multiplicative, therefore
for $x,y\in \mathbb{O}\left( \mathbb{R}\right) ,$ we have 
\begin{equation*}
N\left( xy\right) =N\left( x\right) N\left( y\right) .
\end{equation*}%
\newline
The \textit{real part} of the octonion $x$ is $x_{1}$ and the its \textit{%
vector part} is $%
x_{2}e_{2}+x_{3}e_{3}+x_{4}e_{4}+x_{5}e_{5}+x_{6}e_{6}+x_{7}e_{7}+x_{8}e_{8}%
\in \mathbb{O}\left( \mathbb{R}\right) .\medskip $

In [Co, Sm; 03] p. 55, was described \textit{Hurwitz integers} or \textit{%
Hurwitz Integral Quaternions}, denoted by $\mathcal{H},$ as elements of the
form $q=x_{1}+x_{2}e_{2}+x_{3}e_{3}+x_{4}e_{4}$ where $%
x_{1},x_{2},x_{3},x_{4}$ are in $\mathbb{Z}$ or in $\mathbb{Z}+\frac{1}{2}.$
In the same book, p. 99-105, was \ defined \textit{Octavian Integers} or 
\textit{Octonion Integers} to be the set of the elements spanned by $%
i_{1587},i_{2457},i_{2685},i_{2378}$ over $\mathbb{O}\left( \mathbb{Z}%
\right) ,$ where 
\begin{equation*}
i_{abcd}=\frac{1}{2}\left( e_{a}+e_{b}+e_{c}+e_{d}\right) .
\end{equation*}%
We will denote this ring with $\mathcal{O}.$ $\mathbb{O}\left( \mathbb{Z}%
\right) $ is also called the set of \textit{Gravesian Octonion integers},
the octonions with all coordinates in $\mathbb{Z}$.\medskip

Let $w=\frac{1}{2}\left( 1+\overset{8}{\underset{i=2}{\sum }}e_{i}\right)
\in \mathcal{O},$ be an integer octonion and let $\mathbb{V=\{}a+bw~/~a,b\in 
\mathbb{Z}\mathbb{\}}$. We remark that $N\left( w\right) =2$ and $%
w^{2}-w+2=0.$

Since octonion algebra is a power associative algebra, it results that $%
\mathbb{V}$ is an associative and a commutative ring, $\mathbb{V}\subset $ $%
\mathcal{O}$.\medskip

\textbf{Remark 1.2.} For $x\in \mathbb{V}$\textit{\ } the following
properties are equivalent:

i)\textit{\ }$x$ is invertible in the algebra\textit{\ }$\mathbb{V}.$

ii) $N\left( x\right) =1.$

iii) $x\in \{\pm 1,\pm e_{2}\pm e_{3},\pm e_{4},\pm e_{5},\pm e_{6},\pm
e_{7},\pm e_{8}\}.\medskip $

\textbf{Definition 1.3.} The octonion $x\in \mathbb{V}$ is \textit{prime} in 
$\mathbb{V}$ if $x$ is not an invertible element in $\mathbb{V}$ and if $%
x=ab,$ then $a$ or $b$ is an invertible element in $\mathbb{V}.\medskip $

\textbf{Proposition 1.4. }

\textit{i)} \textit{If} $x,y\in \mathbb{V}$\textit{, then there are} $z,v\in 
\mathbb{V}$ \textit{such that} $x=zy+v,$ \textit{with} $N\left( v\right)
<N\left( y\right) .$

\textit{ii) With the above notation, we have that the remainder} $v$ \textit{%
has the formula}%
\begin{equation}
v=x-\left[ \frac{x\overline{y}}{y\overline{y}}\right] y,  \tag{1.1}
\end{equation}%
\textit{where the symbol} $[~,]$ \textit{is the rounding to the closest
integer. For the octonions, the rounding of an octonion integer can be found
by rounding the coefficients of the basis, separately, to the closest
integer.}\medskip

\textbf{Proof.} i) See [Co, Sm; 03], Theorem 2, p. 109, [Da, Sa, Va; 03]
Proposition 2.6.2, p. 60 and Lemma 2.6.3., p.61.

ii) From the above, we have $v=x-zy$ with $w=x\overline{y}=\overset{8}{%
\underset{i=1}{\sum }}w_{i}e_{i},w_{i}\in \mathbb{Z},$ $w_{i}=az_{i}+r_{i},$ 
$0\leq r_{i}<a,i\in \{1,...,8\}.$ Therefore, $z_{i}=\frac{w_{i}}{a}-\frac{%
r_{i}}{a}=\frac{w_{i}}{y\overline{y}}-\frac{r_{i}}{y\overline{y}}=\frac{(x%
\overline{y})_{i}}{y\overline{y}}-\frac{r_{i}}{y\overline{y}}.$ Rounding to
the closest integer, we have $\left[ \frac{r_{i}}{y\overline{y}}\right] =0,$
since $r_{i}<a.\Box \medskip $

\textbf{Remark 1.5. }Let $x=a+bw\in \mathbb{V}.$ We have that\newline
$N\left( x\right) =x\overline{x}=\left( a+bw\right) \left( a+b\overline{w}%
\right) =a^{2}+ab+2b^{2}=\left( a+\frac{b}{2}\right) ^{2}+7\frac{b^{2}}{4}%
=A^{2}+7B^{2}.\medskip $

\textbf{Proposition 1.6.} [Co; 89],[Sa; 14] \ \textit{Let} $p\in \mathbb{N}$ 
\textit{be a prime number. There are integers} $a,b$ \textit{\ such that} $%
p=a^{2}+ab+2b^{2}$ \textit{if } $p=7k+1,k\in \mathbb{Z}.\Box \medskip $

\textbf{Definition 1.7. } With the above notations, let $\pi =x+yw$ be a
prime integer in $\mathbb{V}$ and $v_{1},v_{2}$ be two elements in $\mathbb{V%
}.$ If there is $v\in \mathbb{V},$~such that $v_{1}-v_{2}=v\pi ,$ then $%
v_{1},v_{2}$ are called \textit{congruent modulo} $\pi $ and it is denoted \ 
$v_{1}\equiv v_{2}$ \textit{mod} $\pi .\medskip $

\textbf{Proposition 1.8. }

\textit{i) The above relation is an equivalence relation on }$\mathbb{V}$%
\textit{. The set of equivalence class is denoted by} $\mathbb{V}_{\pi }$ 
\textit{and is called\ the residue class of } $\mathbb{V}$\textit{\ modulo} $%
\pi .$

\textit{ii) The modulo function }$\mu :\mathbb{V\rightarrow V}_{\pi }$ 
\textit{is} $\mu \left( x\right) =v\ $mod $\pi =x-\left[ \frac{x\overline{y}%
}{y\overline{y}}\right] y,$ where $x=z\pi +v,$ \textit{with} $N\left( \pi
\right) <N\left( y\right) .$

\textit{iii)} $\mathbb{V}_{\pi }$ \textit{is a field } \textit{isomorphic
with} $\mathbb{Z}_{p},p=N(\pi )\medskip ,p$ \textit{a prime number.\medskip }

\textbf{Proof.} i) If $v_{1}\equiv v_{2}$ \textit{mod} $\pi $ and $%
v_{2}\equiv v_{3}$ \textit{mod} $\pi $ then there are $v,v^{\prime }\in 
\mathbb{V}$ such that $v_{1}-v_{2}=v\pi $ and $v_{2}-v_{3}=v^{\prime }\pi .$
It results $v_{1}-v_{3}=(v+v^{\prime })\pi ,$therefore the transitivity. We
will denote with bold the elements from $\mathbb{V}_{\pi }.$

ii) We define $\mathbf{v}_{1}+\mathbf{v}_{2}=\left( v_{1}+v_{2}\right) $%
\textit{mod} $\pi $ and $\mathbf{v}_{1}\cdot \mathbf{v}_{2}=\left(
v_{1}v_{2}\right) $\textit{mod} $\pi .$ These multiplications are well
defined. Indeed, if $v_{1}\equiv v_{1}^{\prime }$ \textit{mod} $\pi $ and $%
v_{2}\equiv v_{2}^{\prime }$ \textit{mod} $\pi ,$ it results $%
v_{1}-v_{1}^{\prime }=u\pi ,v_{2}-v_{2}^{\prime }=u^{\prime }\pi
,u,u^{\prime }\in $ $\mathbb{V},$ therefore $\left( v_{1}+v_{2}\right)
-\left( v_{1}^{\prime }+v_{2}^{\prime }\right) =\left( u+u^{\prime }\right)
\pi .$ From Proposition 1.4 and since $v_{1}=v_{1}^{\prime }+u\pi ,$ $%
v_{2}=v_{2}^{\prime }+u^{\prime }\pi ,~$it results $v_{1}v_{2}=v_{1}^{\prime
}v_{2}^{\prime }+M_{\pi },$ with $M_{\pi }$ a multiple of $\pi .$

Let $f$ \ be the map 
\begin{equation}
f:\mathbb{Z}_{p}\rightarrow \mathbb{V}_{\pi },~f\left( \mathbf{m}\right)
=\mu \left( m+\pi \right) =\left( m+\pi \right) ~\text{\textit{mod }}\pi . 
\tag{1.2}
\end{equation}%
\newline
The map $f$ is well defined, since if $m\equiv m^{\prime }$ \textit{mod} $p$
we have $\left( m+\pi \right) -\left( m^{\prime }+\pi \right) =m-m^{\prime
}=pq=\pi \overline{\pi }q,q\in \mathbb{Z},$ therefore $\left( m+\pi \right)
\equiv \left( m^{\prime }+\pi \right) $ \textit{mod} $\pi .$ \newline

Since $1=v_{1}\pi +v_{2}\overline{\pi },$ if $f\left( \mathbf{m}\right)
=v,v=\left( m+\pi \right) ~$\textit{mod }$\pi \in \mathbb{V}_{\pi },$ we
define $f^{-1}\left( v\right) =m\left( v_{1}\pi \right) +\overline{m}\left(
v_{2}\overline{\pi }\right) .$ Indeed, for $\mathbf{m}\in \mathbb{Z}_{p},$
it results $\mathbf{m=}M\pi +m$ and $\mathbf{m=}\overline{\mathbf{m}}=%
\overline{M}\overline{\pi }+\overline{m}\mathbf{,}$ therefore $m\left(
v_{1}\pi \right) +\overline{m}\left( v_{2}\overline{\pi }\right) =\left( 
\mathbf{m-}M\pi \right) \left( v_{1}\pi \right) +\left( \overline{\mathbf{m}}%
-\overline{M}\overline{\pi }\right) \left( v_{2}\overline{\pi }\right) =%
\mathbf{m}\left( v_{1}\pi +v_{2}\overline{\pi }\right) ~$mod $\pi =\mathbf{m}
$ mod $\pi .$

The map $f$ is a ring morphism. Indeed, $~f\left( \mathbf{m}\right) +f\left( 
\mathbf{m}^{\prime }\right) =\left( m+\pi \right) $\textit{mod} $\pi +\left(
m^{\prime }+\pi \right) $\textit{mod} $\pi =\left( m+m^{\prime }+\pi \right) 
$\textit{mod} $\pi =f\left( \mathbf{m}+\mathbf{m}^{\prime }\right) $ and $%
f\left( \mathbf{m}\right) f\left( \mathbf{m}^{\prime }\right) =\left( m+\pi
\right) \left( m^{\prime }+\pi \right) $\textit{mod}$\pi =$\newline
\textit{=}$\left( mm^{\prime }+\left( m+m^{\prime }\right) \pi +\pi
^{2}\right) $\textit{mod} $\pi =\left( mm^{\prime }+\pi \right) $\textit{mod}
$\pi .$We obtain that $\mathbb{V}_{\pi }$ is isomorphic with $\mathbb{Z}%
_{p}.\Box \medskip $

\textbf{Remark 1.9.} 1) The field $\mathbb{V}_{\pi }$ has the property that
if $x,y\in \mathbb{V}_{\pi }$\textit{,} then there are $z,v\in \mathbb{V}%
_{\pi }$ such that $x=zy+v,$ with $N\left( v\right) <N\left( y\right) $ and $%
\mathbb{V}_{\pi }$ has $N\left( \pi \right) =p$ elements$.$

2) If $\pi =a+bw\in \mathbb{V},$ then in $\mathbb{V}_{\pi }$ two conjugate
elements are not in the same coset. Indeed, if $x=m+nw\in \mathbb{V}_{\pi
},m,n\in \mathbb{Z},$ is in the same coset as $\overline{x}=m-nw,$ therefore 
$\pi ~/~\left( x-\overline{x}\right) ,$ then $\pi ~/~2nw.$ We have $2nw=\pi
q,q\in \mathbb{V},$ then $N\left( \pi \right) ~/~n,$ false, since $N\left(
x\right) =(m+\frac{n}{2})^{2}+7\frac{n^{2}}{4}>N\left( \pi \right)
^{2}>N\left( \pi \right) .\medskip $

\textbf{Remark 1.10. }

1) $\mathbb{O}\left( \mathbb{Z}\right) _{\pi }$ has $N^{4}\left( \pi \right) 
$ elements. ([Ma, Be, Ga; 09]).

2) From Proposition 1.8 and from Remark 1.9, we have that for$%
~v_{i},v_{j}\in \mathbb{V}_{\pi },i,j\in \{1,2,...,p-1\},$ $%
v_{i}+v_{j}=v_{k}~$if and only if $k=i+j$ \textit{mod} $p$ and $v_{i}\cdot
v_{j}=v_{k}~$if and only if $k=i\cdot j$ \textit{mod} $p.$ From here, with
the above notations, we have the following labelling procedure:

1) Let $\pi $ $\in \mathbb{V}$ be a prime, with $N\left( \pi \right) =p,p$ a
prime number, $\pi =a+bw,a,b\in \mathbb{Z}.$

2) Let $s\in \mathbb{Z}$ be the only solution of the equation $a+bx$ \textit{%
mod} $p,~x\in \{0,1,2,...,p-1\}.$

3) The element $k\in \mathbb{Z}_{p}$ is the label of the element $v=m+nw\in 
\mathbb{V}$ if $m+ns=k$ \textit{mod} $p$ and $N\left( v\right) $ is minimum.

The above Remark generalizes to octonions Theorem 1 and Labeling procedure
from \textbf{[}Ne, In,Fa, El, Pa; 01\textbf{].\medskip }

\textbf{Example 1.11.} Using notations from Proposition 1.4, let $p=29$ and $%
\pi =-1+4w,~$with $N\left( \pi \right) =29.$ We compute the field\ $\mathbb{V%
}_{\pi }.$ From the above, we have $\mathbb{U}_{\pi }=\{\left( m+\pi \right)
~$\textit{mod }$\pi $ / $\mathbf{m}\in \mathbb{Z}_{p}\}.$ It results:\newline
$\mathbb{V}_{\pi }=\{0,1,2,3,-3-w,-2-w,-1-w,-w,1-w,2-w,$\newline
$3-w,4-w,-2w-2,2w-2,-2w,-2w+1,-2w+2,$\newline
$2+2w,w-4,w-3,w-2,w-1,w,1+w,2+w,3+w,-3,-2,-1\}.$ Using the above labeling
procedure, we \ have\newline
$f\left( 0\right) =0,f\left( 1\right) =1,f\left( 2\right) =2,f\left(
3\right) =3,~f\left( 4\right) =-3-w,$\newline
$f\left( 5\right) =-2-w,f\left( 6\right) =-1-w,f\left( 7\right) =-w,f\left(
8\right) =1-w,$\newline
$f\left( 9\right) =2-w,f\left( 10\right) =3-w,f\left( 11\right) =4-w,f\left(
12\right) =-2w-2,$\newline
$f\left( 13\right) =2w-2,f\left( 14\right) =-2w,f\left( 15\right)
=-2w+1,f\left( 16\right) =-2w+2,$\newline
$f\left( 17\right) =2+2w,f\left( 18\right) =-4+w,f\left( 19\right)
=-3+w,f\left( 20\right) =-2+w,$\newline
$f\left( 21\right) =-1+w,f\left( 22\right) =w,f\left( 23\right) =1+w,f\left(
24\right) =2+w,$\newline
$f\left( 25\right) =3+w,f\left( 26\right) =-3,f\left( 27\right) =-2,f\left(
28\right) =-1.$

\begin{equation*}
\end{equation*}

\bigskip \textbf{2.} \textbf{Codes over } $\mathbb{V}_{\pi }$

\begin{equation*}
\end{equation*}

Generalizing the Hurwitz weight from [Gu; 13], we define the \textit{%
Cyley-Dickson weight}, denoted $d_{C}.$ Let $\pi $ be a prime in $\mathbb{V}%
, $ $\mathbb{\pi =}a+bw.$ Let $x\in \mathbb{V},$ $x=a_{0}+b_{0}w.$ The 
\textit{Cyley-Dickson weight} of $x$ is defined as $w_{C}\left( x\right)
=\left\vert a_{0}\right\vert +\left\vert b_{0}\right\vert ,$where $%
x=a_{0}+b_{0}w$ \textit{mod}$~\pi ,$ with $\left\vert a_{0}\right\vert
+\left\vert b_{0}\right\vert $ minimum. \newline

The \textit{Cyley-Dickson distance }between $x,y\in \mathbb{V}_{\pi }$ is
defined as 
\begin{equation*}
d_{C}\left( x,y\right) =w_{C}\left( x-y\right) .
\end{equation*}
We will prove that $d_{C}$ is a metric. Indeed, for $x,y,z$ three elements
in $\mathbb{V}_{\pi },$ we have $d_{C}\left( x,y\right) =w_{C}\left( \alpha
_{1}\right) =\left\vert a_{1}\right\vert +\left\vert b_{1}\right\vert ,$
where $\alpha _{1}=x-y=a_{1}+b_{1}$ \textit{mod} $\pi $ is an element in $%
\mathbb{V}_{\pi }$ and $\left\vert a_{1}\right\vert +\left\vert
b_{1}\right\vert $ is minimum.\newline
$d_{C}\left( y,z\right) =w_{C}\left( \alpha _{2}\right) =\left\vert
a_{2}\right\vert +\left\vert b_{2}\right\vert ,$where $\alpha
_{2}=x-y=a_{2}+b_{2}$ \textit{mod} $\pi $ is an element in $\mathbb{V}_{\pi
} $ and $\left\vert a_{2}\right\vert +\left\vert b_{2}\right\vert $ is
minimum.\newline
$d_{C}\left( x,z\right) =w_{C}\left( \alpha _{3}\right) =\left\vert
a_{3}\right\vert +\left\vert b_{3}\right\vert ,$where $\alpha
_{3}=x-y=a_{3}+b_{3}$ \textit{mod} $\pi $ is an element in $\mathbb{V}_{\pi
} $ and $\left\vert a_{3}\right\vert +\left\vert b_{3}\right\vert $ is
minimum.\newline
We have $x-y=\alpha _{2}+\alpha _{3}$ \textit{mod} $\pi .$ It results $%
w_{C}\left( \alpha _{2}+\alpha _{3}\right) \geq w_{C}\left( \alpha
_{1}\right) $ since $w_{C}\left( \alpha _{1}\right) =\left\vert
a_{1}\right\vert +\left\vert b_{1}\right\vert $ is minimum.\medskip

\textbf{Remark 2.1.} The maximum Cayley-Dickson distance $d_{C_{\max }}$%
between any two elements from $\mathbb{V}_{\pi }$ has the property that $%
d_{C_{\max }}\leq \left\vert a\right\vert +\left\vert b\right\vert ,$ with $%
\pi =a+bw$.\medskip

\textbf{Remark 2.2.} Since the Octonion algebra is alternative, due to
Artin's Theorem (see [Sc; 66]), each two nonzero different elements generate
an associative algebra. From here, for $x,y\in \mathbb{O}$ we have that $%
x^{m}\left( x^{n}y\right) =x^{m+n}y,$ for all $m,n\in \mathbb{Z}.\medskip $

In the following, we will consider $\pi $ a prime in $\mathbb{V}$ of the
norm $N\left( \pi \right) =\pi \overline{\pi }=p,p=7l+1,l\in \mathbb{Z}.$
Let $\alpha _{1},\alpha _{2}$ be two primitive elements (of order $p-1)$ in $%
\mathbb{V}_{\pi },$ such that $\alpha _{1}^{\frac{p-1}{7}}=w,\alpha _{2}^{%
\frac{p-1}{7}}=-w.$ Let $\alpha \in \{\alpha _{1},\alpha _{2}\}.$ We will
consider codes of length $n=\frac{p-1}{7}.$

Let $C$ be the code defined by the parity-check matrix $H,$%
\begin{equation}
H=\left( 
\begin{array}{ccccc}
1 & \alpha & \alpha ^{2} & ... & \alpha ^{n-1} \\ 
1 & \alpha ^{8} & \alpha ^{16} & ... & \alpha ^{8(n-1)} \\ 
... & ... & ... & ... & ... \\ 
1 & \alpha ^{7k+1} & \alpha ^{2\left( 7k+1\right) } & ... & \alpha
^{(7k+1)\left( n-1\right) }%
\end{array}%
\right) ,  \tag{2.1}
\end{equation}%
with $k<n.$ We know that $c$ is a codeword in $C$ if and only if $Hc^{T}=0.$
From here, if \ we consider the associate code polynomial $c\left( x\right) =%
\underset{i=0}{\overset{n-1}{\sum }}c_{i}x^{i},$ we have that $c\left(
\alpha ^{7l+1}\right) =0,l\in \{0,1,...,k\}.$ We consider the polynomial $%
g\left( x\right) =\left( x-\alpha \right) \left( x-\alpha ^{8}\right)
...\left( x-\alpha ^{7k+1}\right) .$ Since the elements $\alpha ,\alpha
^{8},...,\alpha ^{7k+1}$ are distinct, from [Li, Xi; 04], Lemma 8.1.6, we
have that $c\left( x\right) $ is divisible by the generator polynomial $%
g\left( x\right) .$ Since $g\left( x\right) ~/~($ $x^{n}\pm w)$, therefore $%
g\left( x\right) $ is the generator polynomial for the code $C.$ It results
that $C$ is a principal ideal in the ring $\mathbb{V}_{\pi }~/$ $\left(
(x^{n}\pm w)\right) .\medskip $

\textbf{Theorem 2.3.} \textit{Let } $C$ \textit{be the code defined on }$%
\mathbb{V}_{\pi }$\textit{\ by the parity check matrix }%
\begin{equation}
H=\left( 
\begin{array}{ccccc}
1 & \alpha & \alpha ^{2} & ... & \alpha ^{n-1}%
\end{array}%
\right) .  \tag{2.2}
\end{equation}%
\textit{The code} $C$ \textit{is able to correct all errors pattern of the
form} $e\left( x\right) =e_{t}x^{t},$ \textit{with} $0\leq w_{C}\left(
e_{t}\right) \leq 1$ \textit{and any errors pattern of the form }$e\left(
x\right) =e_{t}x^{t},$ \textit{with} $w_{C}\left( e_{t}\right) =3,e_{t}=\pm
w^{2}.$ $\medskip $

\textbf{Proof.} Let $r\left( x\right) =c\left( x\right) +e\left( x\right) $
be the received polynomial, with $c(x)$ the codeword polynomial and $%
e(x)=e_{t}x^{t}$ denotes the error polynomial with $0\leq w_{C}\left(
e_{t}\right) \leq 1.$ Since $\alpha ^{n}=w,$ or $\alpha ^{n}=-w$ and $%
w^{2}=w-2,d_{C}\left( w^{2}\right) =3,$ it results $e_{t}=\alpha ^{nl}.$ We
have the syndrome $s_{1}=\alpha ^{t+nl}=\alpha ^{L},$ with $t,L\in \mathbb{Z}%
,0\leq t,L\leq n-1.$ If we reduce $L$ modulo $n,$ we obtain $t,$ the
location of the error, and from here, $l=\frac{L-t}{n},~$the value of the
error.$\Box \medskip $

\textbf{Example 2.4.} With the above notation, let $\pi
=7+2w,p=71,n=10,\alpha =2-2w,w=\alpha ^{10}$ and the parity check matrix 
\begin{equation*}
H=\left( 
\begin{array}{cccccccccc}
1 & \alpha & \alpha ^{2} & \alpha ^{3} & \alpha ^{4} & \alpha ^{5} & \alpha
^{6} & \alpha ^{7} & \alpha ^{8} & \alpha ^{9}%
\end{array}%
\right) .
\end{equation*}%
Supposing that the received vector is $r=\left( w,1,w-1,1,1,0,0,0,1,1\right)
.$ Using MAPLE software, we compute the syndrome. It results $%
S=Hr^{t}=-5-2w=\alpha ^{14}~$\textit{mod} $\pi .$ We get $L=14,$ therefore
the location of the error is $l=L$ \textit{mod} $14=4$ \textit{mod} $10$.
The value is $w=\alpha ^{14-4}=\alpha ^{10}$ \textit{mod} $\pi .$ Therefore
the corrected vector is \newline
$c$=$r-\left( 0,0,0,0,w,0,0,0,0,0\right) $=$\left(
w,1,w-1,1,1-w,0,0,0,1,1\right) $ \textit{mod} $\pi .\medskip $

\textbf{Theorem 2.5.} \textit{Let }$\ C$ \textit{be a code defined by the
parity-check matrix} \newline
\begin{equation}
H=\left( 
\begin{array}{ccccc}
1 & \alpha & \alpha ^{2} & ... & \alpha ^{n-1} \\ 
1 & \alpha ^{8} & \alpha ^{16} & ... & \alpha ^{8(n-1)}%
\end{array}%
\right) .  \tag{2.3}
\end{equation}%
\textit{Then} $C$ \textit{can corrects any error pattern of the form} $%
e\left( x\right) =e_{i}x^{i},$ $0\leq i\leq n-1,$ \textit{with}$~e_{i}\in 
\mathbb{V}_{\pi }.\medskip $

\textbf{Proof.} Let $r\left( x\right) =c\left( x\right) +e\left( x\right) $
be the received polynomial, with $c(x)$ the codeword polynomial and $%
e(x)=e_{i}x^{i}$ denotes the error polynomial with $e_{i}\in \mathbb{V}_{\pi
}.$ Then the corresponding vector of the polynomial $r(x)$ is $r=c+e$ and we
will compute the syndrome $S$ of $r.$ We have $e_{i}=\alpha ^{q},0\leq q\leq
7n-1.$ Therefore the syndrome is 
\begin{equation*}
S\text{=}Hr^{t}\text{=}\left( 
\begin{array}{c}
s_{1}=\alpha ^{i+q}=\alpha ^{M_{1}} \\ 
s_{8}=\alpha ^{8i+q}=\alpha ^{M_{2}}%
\end{array}%
\right) .
\end{equation*}%
We obtain $a^{i+q-M_{1}}=1,$ with $i+q=M_{1}$ \textit{mod}$(p-1)$ and $%
\alpha ^{8i+q-M_{2}}=1,$ with $8i+q=M_{2}$ \textit{mod}$(p-1).$ We get $%
7i=(M_{2}-M_{1})$ \textit{mod}$(p-1),$ therefore the unique solution of the
system is $i=\frac{M_{2}-M_{1}}{7}$ \textit{mod~}$n$ and $q=(M_{1}-i)$ 
\textit{mod}$(p-1).$ In this way, we found the location and the value of the
error.$\Box \medskip $

\textbf{Example 2.6.} Let $\pi =-1+4w,p=29,n=4,\alpha =1-w,-w=\alpha ^{4}$ 
\textit{mod} $\pi ,$ and the parity check matrix%
\begin{equation*}
H=\left( 
\begin{array}{cccc}
1 & \alpha & \alpha ^{2} & \alpha ^{3} \\ 
1 & \alpha ^{8} & \alpha ^{16} & \alpha ^{24}%
\end{array}%
\right) .
\end{equation*}%
Supposing that the received vector is $r=\left( \alpha ,\alpha ^{2},1,\alpha
^{3}\right) =\left( 1-w,2-2w,1,-3+w\right) .$Using MAPLE software, we
compute the syndrome. It results 
\begin{equation*}
S=Hr^{t}=\left( 
\begin{array}{c}
s_{1}=\alpha ^{7} \\ 
s_{8}=\alpha ^{7}%
\end{array}%
\right) .
\end{equation*}%
The location of the error is $i=\frac{7-7}{7}=0$ \textit{mod} $4$ and the
value of the error is $\alpha ^{7-0}=\alpha ^{7}=17=(2+2w)$ \textit{mod} $%
\pi .$ Therefore the corrected vector is\newline
$c=r-\left( 2+2w,0,0,0\right) =\left( -1-3w,2-2w,1,-3+w\right) $\textit{mod~}%
$\pi =$\newline
$=\left( w-2,2-2w,1,-3+w\right) .\medskip $

\textbf{Theorem 2.7.} \textit{Let }$\ C$ \textit{be a code defined by the
parity-check matrix} \newline
\begin{equation}
H=\left( 
\begin{array}{ccccc}
1 & \alpha & \alpha ^{2} & ... & \alpha ^{n-1} \\ 
1 & \alpha ^{8} & \alpha ^{16} & ... & \alpha ^{8(n-1)} \\ 
1 & \alpha ^{15} & \alpha ^{30} & ... & \alpha ^{15(n-1)}%
\end{array}%
\right) .  \tag{2.4}
\end{equation}%
\textit{Then} $C$ \textit{can find location and/or can correct some errors
pattern of the form} $e\left( x\right) =e_{i}x^{i},$ $0\leq i\leq n-1,$ 
\textit{with}$~e_{i}\in \mathbb{V}_{\pi }.\medskip $

\textbf{Proof.} Using notations from the above Theorem, we have $%
e_{i}=\alpha ^{q},0\leq q\leq 7n-1.$ Therefore the syndrome is: 
\begin{equation*}
S\text{=}Hr^{t}\text{=}\left( 
\begin{array}{c}
s_{1}=\alpha ^{i+q}=\alpha ^{M_{1}} \\ 
s_{8}=\alpha ^{8i+q}=\alpha ^{M_{2}} \\ 
s_{15}=\alpha ^{15i+q}=\alpha ^{M_{3}}%
\end{array}%
\right) .
\end{equation*}

Since the rank of the matrix $\left( 2.4\right) $ is $3,$ then this system
has always solution. We obtain $a^{i+q-M_{1}}=1,$ with $i+q=M_{1}$ \textit{%
mod}$(p-1),$ $\alpha ^{8i+q-M_{2}}=1,$ with $8i+q=M_{2}$ \textit{mod}$%
(p-1),~\alpha ^{15i+q-M_{3}}=1,~$with $15i+q=M_{3}~$\textit{mod}$(p-1).$ We
can find the location of the error if $7i=(M_{2}-M_{1})$ \textit{mod}$(p-1),$
$7i=(M_{3}-M_{2})$ \textit{mod}$(p-1),$ or, equivalently, \newline
$i=\frac{M_{2}-M_{1}}{7}$ \textit{mod~}$n=\frac{M_{3}-M_{2}}{7}$ \textit{mod~%
}$n$ and the value of the error $q$ if \newline
$(M_{1}-i)$ \textit{mod}$(p-1)=(M_{2}-8i)$ \textit{mod}$(p-1)=(M_{3}-15i)$ 
\textit{mod}$(p-1)(=q).$ $\Box \medskip $

\textbf{Example 2.8.\medskip }

1) Let $\pi =-1+4w,p=29,n=4,\alpha =1-w,-w=\alpha ^{4}$ \textit{mod} $\pi ,$
and the parity check matrix 
\begin{equation*}
H=\left( 
\begin{array}{cccc}
1 & \alpha & \alpha ^{2} & \alpha ^{3} \\ 
1 & \alpha ^{8} & \alpha ^{16} & \alpha ^{24} \\ 
1 & \alpha ^{15} & \alpha ^{30} & \alpha ^{45}%
\end{array}%
\right) .
\end{equation*}%
Supposing that the received vector is $r=\left( 1-w,2-2w,1,-3+w\right) .$
Using MAPLE software, we compute the syndrome. It results 
\begin{equation*}
S=Hr^{t}=\left( 
\begin{array}{c}
s_{1}=\alpha ^{7}=\alpha ^{i+q} \\ 
s_{8}=\alpha ^{7}=\alpha ^{8i+q} \\ 
s_{15}=\alpha ^{27}=\alpha ^{15i+q}%
\end{array}%
\right) .
\end{equation*}%
The location of the error is $i=\frac{7-7}{7}=\frac{27-7}{7}=0$ \textit{mod} 
$4.$ We can't find the value of the error since $\alpha ^{7-0}=\alpha
^{7}=17=(2+2w)$ \textit{mod} $\pi $ is different from $\alpha ^{27-0}=\alpha
^{27}=11=(4-w)$ \textit{mod} $\pi .$

2) In the same conditions, supposing that the received vector is\newline
$r=\left( 1,\alpha ^{3},1,\alpha ^{2}\right) =\left( 1,-3+w,1,-1-w\right) .$
Again, using MAPLE, the syndrome is 
\begin{equation*}
S=Hr^{t}=\left( 
\begin{array}{c}
s_{1}=\alpha ^{21}=\alpha ^{i+q} \\ 
s_{8}=\alpha ^{11}=\alpha ^{8i+q} \\ 
s_{15}=\alpha ^{18}=\alpha ^{15i+q}%
\end{array}%
\right) .
\end{equation*}%
We can't find the location and the value of the error, since $2=\frac{11-21}{%
7}$ \textit{mod} $4\neq \frac{18-11}{7}$ \textit{mod} $\pi =1.$

3) If we suppose that the received vector is $r=\left( 5,0,0,0\right)
=\left( -2-w,0,0,0\right) ,$ the syndrome is%
\begin{equation*}
S=Hr^{t}=\left( 
\begin{array}{c}
s_{1}=\alpha ^{26} \\ 
s_{8}=\alpha ^{26} \\ 
s_{15}=\alpha ^{26}%
\end{array}%
\right) .
\end{equation*}%
The location of the error is $0$ and the value of the error is $5.$
Therefore the corrected vector is $\left( 0,0,0,0\right) .\medskip $

\textbf{Theorem 2.9.} \textit{Let }$\ C$ \textit{be a code defined by the
parity-check matrix} \newline
\begin{equation}
H=\left( 
\begin{array}{ccccc}
1 & \alpha & \alpha ^{2} & ... & \alpha ^{n-1} \\ 
1 & \alpha ^{8} & \alpha ^{16} & ... & \alpha ^{8(n-1)} \\ 
1 & \alpha ^{15} & \alpha ^{30} & ... & \alpha ^{15(n-1)} \\ 
1 & \alpha ^{22} & \alpha ^{44} & ... & \alpha ^{22\left( n-1\right) }%
\end{array}%
\right) .  \tag{2.5}
\end{equation}%
\textit{Then} $C$ \textit{can correct some errors pattern of the form} $%
e\left( x\right) =e_{i}x^{i}+e_{j}x^{j},$ $0\leq i,j\leq n-1,$ \textit{with}$%
~e_{i},e_{j}\in \mathbb{V}_{\pi }.\medskip $

\textbf{Proof. }\ We will prove the general case, when we have two errors.
We have $e_{i}=\alpha ^{q}\neq 0$ and $e_{j}=\alpha ^{t}\neq 0,q,t\in 
\mathbb{Z}$. We obtain the syndrome:%
\begin{equation*}
S\text{=}Hr^{t}\text{=}\left( 
\begin{array}{c}
s_{1}=\alpha ^{i+q}+\alpha ^{j+t} \\ 
s_{8}=\alpha ^{8i+q}+\alpha ^{8j+t} \\ 
s_{15}=\alpha ^{15i+q}+\alpha ^{15j+t} \\ 
s_{22}=\alpha ^{22i+q}+\alpha ^{22j+t}%
\end{array}%
\right) .
\end{equation*}%
Denoting $\alpha ^{i+q}=A$ and $\alpha ^{j+t}=B,$ it results

\begin{equation}
S\text{=}Hr^{t}\text{=}\left( 
\begin{array}{c}
s_{1}=A+B \\ 
s_{8}=\alpha ^{8i}A+\alpha ^{8j}B \\ 
s_{15}=\alpha ^{15i}A+\alpha ^{15j}B \\ 
s_{22}=\alpha ^{22i}A+\alpha ^{22j}B%
\end{array}%
\right)  \tag{2.6}
\end{equation}

If the system $\left( 2.6\right) $ admits only one solution, then the code $%
C $ can correct two errors. First, we will prove the following Lemma.\medskip

\textbf{Lemma.} \ \textit{With the above notations, if we have two errors,
we obtain} $\alpha ^{7i}\neq \alpha ^{7j},0\leq i,j\leq n-1$ \textit{and} $%
s_{1}s_{15}\neq s_{8}^{2}.\medskip $

\textit{Proof.} If \ $\alpha ^{7i}=\alpha ^{7j},$then $\alpha ^{7(i-j)}=1$
and $7n~/~7(i-j),$ false. Supposing that $s_{1}s_{15}-s_{8}^{2}=0,$ we have $%
s_{1}s_{15}=s_{8}^{2}.$ $\ $For $x=$ $\alpha ^{i+q},$ it results $\alpha
^{14i}s_{1}x+\alpha ^{14j}s_{1}^{2}-\alpha ^{14j}s_{1}x=\left( \alpha
^{7i}-\alpha ^{7j}\right) ^{2}x^{2}+\alpha ^{14j}s_{1}^{2}+2\alpha
^{7j}\left( \alpha ^{7i}-\alpha ^{7j}\right) s_{1}x.$ We get $\left( \alpha
^{7i}-\alpha ^{7j}\right) ^{2}x^{2}+2\alpha ^{7i+7j}s_{1}x-\alpha
^{14i}s_{1}x-\alpha ^{14j}s_{1}x=0.$ From here, $x=0$ or $x=\frac{-2\alpha
^{7i+7j}s_{1}+\alpha ^{14i}s_{1}+\alpha ^{14j}s_{1}}{\left( \alpha
^{7i}-\alpha ^{7j}\right) ^{2}}=s_{1}.$ If $x=$ $\alpha ^{i+q}=0,$ false. If 
$x=$ $\alpha ^{i+q}=s_{1}$ implies $\alpha ^{j+t}=0,$false.\medskip

Now, we return to the proof of the Theorem and we are under conditions $%
\alpha ^{7i}\neq \alpha ^{7j},0\leq i,j\leq n-1$ and $s_{1}s_{15}\neq
s_{8}^{2}.\medskip $ For $B=s_{1}-A,$ it results:\newline
$A\left( \alpha ^{7i}-\alpha ^{7j}\right) =s_{8}-s_{1}\alpha ^{7j}$\newline
$A\left( \alpha ^{14i}-\alpha ^{14j}\right) =s_{15}-s_{1}\alpha ^{14j}$%
\newline
$A\left( \alpha ^{21i}-\alpha ^{21j}\right) =s_{22}-s_{1}\alpha ^{21j}$. We
obtain: 
\begin{equation*}
s_{15}-s_{1}\alpha ^{14j}=\left( s_{8}-s_{1}\alpha ^{7j}\right) \left(
\alpha ^{7i}+\alpha ^{7j}\right)
\end{equation*}%
and%
\begin{equation*}
s_{22}-s_{1}\alpha ^{21j}=\left( s_{8}-s_{1}\alpha ^{7j}\right) \left(
\alpha ^{14i}+\alpha ^{7i}\alpha ^{7j}+\alpha ^{14j}\right) .
\end{equation*}%
\newline
Denoting $\alpha ^{7i}+\alpha ^{7j}=s_{7}$ and $\alpha ^{7i}\alpha
^{7j}=p_{7},$ we have

\begin{equation*}
s_{15}-s_{8}s_{7}+p_{7}s_{1}=0
\end{equation*}%
and 
\begin{equation*}
\left( s_{8}-s_{1}\alpha ^{7j}\right) \left( s_{7}^{2}-p_{7}\right)
=s_{22}-s_{1}\alpha ^{21j}.
\end{equation*}%
It results%
\begin{equation*}
p_{7}=\frac{s_{8}s_{7}-s_{15}}{s_{1}}
\end{equation*}%
and%
\begin{equation*}
s_{7}(s_{1}s_{15}-s_{8}^{2})=s_{1}s_{22}-s_{8}s_{15}.
\end{equation*}

We obtain 
\begin{equation*}
s_{7}=\frac{s_{1}s_{22}-s_{8}s_{15}}{s_{1}s_{15}-s_{8}^{2}}
\end{equation*}%
and for $p_{7}$ we get 
\begin{equation*}
p_{7}=\frac{s_{8}s_{22}-s_{15}^{2}}{s_{1}s_{15}-s_{8}^{2}}.
\end{equation*}

From here, solving the equation $x^{2}-s_{7}x+p_{7}=0,$ we will find the
locations and the values of the errors. $\Box \medskip $\newline

\textbf{Example 2.\medskip 10.}

1) Let $\pi =-1+4w,p=29,n=4,\alpha =1-w,-w=\alpha ^{4}$ \textit{mod} $\pi ,$
and the parity check matrix%
\begin{equation*}
H=\left( 
\begin{array}{cccc}
1 & \alpha & \alpha ^{2} & \alpha ^{3} \\ 
1 & \alpha ^{8} & \alpha ^{16} & \alpha ^{24} \\ 
1 & \alpha ^{15} & \alpha ^{30} & \alpha ^{45} \\ 
1 & \alpha ^{22} & \alpha ^{44} & \alpha ^{66}%
\end{array}%
\right) .
\end{equation*}%
Supposing that that the received vector is\newline
$r=\left( 1,\alpha ^{3},1,\alpha ^{2}\right) =\left( 1,-3+w,1,-1-w\right) .$
Using again MAPLE, the syndrome is%
\begin{equation*}
S\text{=}Hr^{t}\text{=}\left( 
\begin{array}{c}
s_{1}=\alpha ^{21} \\ 
s_{8}=\alpha ^{11} \\ 
s_{15}=\alpha ^{18} \\ 
s_{22}=\alpha ^{20}%
\end{array}%
\right) .
\end{equation*}%
We obtain 
\begin{equation*}
s_{7}=\frac{s_{1}s_{22}-s_{8}s_{15}}{s_{1}s_{15}-s_{8}^{2}}=(2+w)\text{ 
\textit{mod} }\pi \text{ }
\end{equation*}%
and

\begin{equation*}
p_{7}=\frac{s_{8}s_{22}-s_{15}^{2}}{s_{1}s_{15}-s_{8}^{2}}=\text{ }1\text{ 
\textit{mod} }\pi \text{ }.
\end{equation*}

Equation $x^{2}-(2+w)x+1=0$ has no roots in $\mathbb{V}_{\pi },$ therefore
we can't find the locations and the values of the errors$.$

2) If the received vector is $r=\left( 5,0,1,0\right) =\left(
-2-w,0,1,0\right) ,$ the syndrome is%
\begin{equation*}
S\text{=}Hr^{t}\text{=}\left( 
\begin{array}{c}
s_{1}=\alpha ^{27}=11 \\ 
s_{8}=\alpha ^{14}=28 \\ 
s_{15}=\alpha ^{27}=11 \\ 
s_{22}=\alpha ^{14}=28%
\end{array}%
\right) .
\end{equation*}%
We get $s_{7}=0$ and $p_{7}=-1$ and $\alpha ^{7i}=1,\alpha ^{7j}=28$ mod $%
\pi .$ It results $\alpha ^{i}=1,\alpha ^{j}=4=\alpha ^{10},$ then $i=0$ and 
$j=10$ mod $4=2.$ The errors are on the positions $0$ and $2.$ The corrected
vector are $c=\left( 4,0-3,0\right) =\left( -3-w,0,-3,0\right) .$

\begin{equation*}
\end{equation*}

\textbf{Remark 2.11.} Above Theorems generalized to octonions Theorems
7,8,9,10,11,13,14,15 from [Ne, In,Fa, El, Pa; 01] and Theorems 4,5,6,7 from
[Gu; 13].\medskip

\textbf{Remark 2.12.} In this situation, when $p=7k+1,k=6l,l\in \mathbb{Z},$
and when the considered alphabets have the same cardinality, the code rate
of the codes defined on $\mathbb{V}_{\pi }$ is grater than the codes defined
in [Gu; 13] on $\mathcal{R}_{\pi }$, but smaller than the codes defined on $%
\mathcal{H}_{\pi }.$ Here $\mathcal{H}$ is the set of all Hurwitz integers, $%
\mathcal{R}=\{a+bw,a,b\in \mathbb{Z}\},w=\frac{1}{2}\left( 1+i+j+k\right) ,$
with $\{1,i,j,k\}$ a basis in the Quaternion algebra and $\mathcal{R}_{\pi
}, $ $\mathcal{H}_{\pi }$ is the quotient rings modulo $\pi ,$ $\pi $ a
prime quaternion. If $C_{1}$ is a code over $\mathcal{R}_{\pi }$ of length $%
n_{1}=\frac{p-1}{6},$ $C_{2}$ a code over $\mathbb{V}_{\pi }$ of length $%
n_{2}=\frac{p-1}{28},$ with $N\left( \pi \right) =p.$ If $C_{1},C_{2}$ have
the same dimension $k,$ we obtain that the rate $R_{C_{2}}$ of the code $%
C_{2}$ is always greater than the rate $R_{C_{1}}$ of the code $C_{1}.$
Indeed, $R_{C_{2}}=\frac{7k}{p-1}$ and $R_{C1}=\frac{6k}{p-1}.\medskip $

\medskip

\textbf{Conclusion.} In this paper we defined codes over a subset of
Octonion Integer and we gave decoding algorithms for these codes. Moreover,
comparing these codes with the codes defined on Hurwitz integers, we remark
that the code rate in the first case is better.

\begin{equation*}
\end{equation*}

\textbf{References}%
\begin{equation*}
\end{equation*}%
\newline
[Ba; 09] J. W. Bales, \textit{A Tree for Computing the Cayley-Dickson Twist}%
, Missouri J. Math. Sci., \textbf{21(2)(2009)}, 83--93.\medskip \newline
[Co, Sm; 03] J.H. Conway, D.A. Smith, \textit{On Quaternions and Octonions},
A.K. Peters, Natick, Massachusetts, 2003.\medskip \newline
[Co; 89] D. Cox, \textit{Primes of the Form} $x^{2}+ny^{2}$\textit{: Fermat,
Class Field Theory and Complex Multiplication}, A Wiley - Interscience
Publication, New York, 1989.\medskip \newline
[Da, Sa, Va; 03] G. Davidoff, P. Sarnak, A. Valette, \textit{Elementary
Number Theory, Group Theory, and Ramanujan Graphs}, Cambridge University
Pres, 2003.\medskip \newline
[Gh, Fr; 10] F. Ghaboussi, J. Freudenberger, \textit{Codes over Gaussian
integer rings}, 18th Telecommunications forum TELFOR 2010, 662-665.\medskip\ 
\newline
[Sa, Na, Re, Sa; 13] M. Gonz\'{a}lez Sarabia, J. Nava Lara, C. Renter\'{\i}a
M\'{a}rquez, E. Sarmiento Rosales, \textit{Parameterized Codes over Cycles,}
An. \c{S}t. Univ. Ovidius Constan\c{t}a, \textbf{21(3)(2013)},
241-255.\medskip\ \newline
[Gu; 13] M. G\"{u}zeltepe, \textit{Codes over Hurwitz integers}, Discrete
Math, \textbf{313(5)(2013)}, 704-714.\medskip \newline
[Hu; 94] K. Huber, \textit{Codes over Gaussian integers}, IEEE Trans.
Inform. Theory, \textbf{40(1994), } 207--216.\medskip \newline
[Ko, Mo, Ii, Ha, Ma;10] H. Kostadinov, H. Morita, N. Iijima, A. J. Han
Vinck, N. Manev, S\textit{oft Decoding Of Integer Codes and Their
Application to Coded Modulation}, IEICE Trans. Fundamentals, \textbf{%
E39A(7)(2010)}, 1363-1370.\medskip \newline
[Li, Xi; 04] S. Ling, C. Xing, \textit{Coding Theory A First Course},
Cambridge University Press, 2004.\medskip \newline
[Ma, Be, Ga; 09] C. Martinez, R. Beivide, E. Gabidulin, \textit{Perfect
codes from Cayley graphs over Lipschitz integers}, IEEE Trans. Inform.
Theory \textbf{55(8)(2009),} 3552--3562.\medskip \newline
[Mo, Ha, Ko; 04] H. Morita, A. J. Han Vinck, H. Kostadinov, \textit{On Soft
Decoding of Coded QAM Using Integer Codes}, International Symposium on
Information Theory and its Applications, ISITA2004, Parma, Italy,
1321-1325.\medskip\ \newline
[Ne, In,Fa, El, Pa; 01] T.P. da N. Neto, J.C. Interlando, M.O. Favareto, M.
Elia, R. Palazzo Jr., \textit{Lattice constellation and codes from quadratic
number fields}, IEEE Trans. Inform.Theory \textbf{47(4)(2001)}
1514--1527.\medskip \newline
[Ni, Hi; 08] S. Nishimura, T. Hiramatsu, \textit{A generalization of the Lee
distance and error correcting codes}, Discrete Appl Math, \textbf{156(2008)}%
, 588 -- 595.\medskip \newline
[Ri; 95] J. Rif\`{a}, \textit{Groups of Complex Integer Used as QAM Signals}%
, IEEE Transactions on Information Theory, \textbf{41(5)(1995)},
1512-1517.\medskip \newline
[Sa; 14] D. Savin, \textit{Some central simple algebras which split}, will
appear in An. \c{S}t. Univ. Ovidius Constan\c{t}a.\medskip \newline
[Sc; 66] R.D. Schafer, \textit{An Introduction to Nonassociative Algebras},
Academic Press, New York, 1966.%
\begin{equation*}
\end{equation*}

Cristina FLAUT

{\small Faculty of Mathematics and Computer Science,}

{\small Ovidius University,}

{\small Bd. Mamaia 124, 900527, CONSTANTA,}

{\small ROMANIA}

{\small http://cristinaflaut.wikispaces.com/}

{\small http://www.univ-ovidius.ro/math/}

{\small e-mail:}

{\small cflaut@univ-ovidius.ro}

{\small cristina\_flaut@yahoo.com}

\end{document}